\RequirePackage[pagewise,mathlines]{lineno}

\documentclass[twocolumn,aps,showpacs,superscriptaddress]{revtex4-1}
\usepackage{savesym}
\usepackage{amssymb}
\usepackage{xcolor}
\usepackage{bbding}
\usepackage{multirow}
\usepackage{graphicx}

\usepackage{dcolumn}
\restoresymbol{TXT}{affil}
\usepackage{bm}

\begin{document}

\def\Journal#1#2#3#4{{#1} {\bf #2}, #3 (#4)}

\def\NCA{Nuovo Cimento}
\def\NIM{Nucl. Instr. Meth.}
\def\NIMA{{Nucl. Instr. Meth.} A}
\def\NPB{{Nucl. Phys.} B}
\def\NPA{{Nucl. Phys.} A}
\def\PLB{{Phys. Lett.}  B}
\def\PRL{Phys. Rev. Lett.}
\def\PRC{{Phys. Rev.} C}
\def\PRD{{Phys. Rev.} D}
\def\ZPC{{Z. Phys.} C}
\def\JPG{{J. Phys.} G}
\def\CPC{Comput. Phys. Commun.}
\def\EPJ{{Eur. Phys. J.} C}
\def\PR{Phys. Rept.}
\def\PRV{Phys. Rev.}
\def\JHEP{JHEP}

\preprint{}
\title{Systematic study of the experimental measurements on $J/\psi$ cross section and kinematic distribution in $p+p$ collisions at different energies}
\date{\today}
\author{Wangmei Zha}\address{University of Science and Technology of China, Hefei, China}\address{Brookhaven National Laboratory, New York, USA}
\author{Bingchu Huang}\address{Brookhaven National Laboratory, New York, USA}
\author{Rongrong Ma}\address{Brookhaven National Laboratory, New York, USA}
\author{Lijuan Ruan}\address{Brookhaven National Laboratory, New York, USA}
\author{Zebo Tang}\email{zbtang@ustc.edu.cn}\address{University of Science and Technology of China, Hefei, China}
\author{Zhangbu Xu}\address{Brookhaven National Laboratory, New York, USA}
\author{Chi Yang}\address{University of Science and Technology of China, Hefei, China}\address{Brookhaven National Laboratory, New York, USA}
\author{Qian Yang}\address{University of Science and Technology of China, Hefei, China}\address{Brookhaven National Laboratory, New York, USA}
\author{Shuai Yang}\address{University of Science and Technology of China, Hefei, China}\address{Brookhaven National Laboratory, New York, USA}

\begin{abstract}
The world experimental data on cross section and kinematic distribution in $p+p$ and $p+A$ collisions at $\sqrt{s}$ = 6.8 - 7000 GeV are examined in systematic way. The  $\sqrt{s}$ dependence of the inclusive cross section, rapidity and transverse momentum distributions are studied phenomenologically. We explore empirical formulas to obtain the total cross section, rapidity and transverse momentum ($p_{T}$) distribution.  This is crucial for the interpretation of A$+$A $J/\psi$ results at RHIC when the $p+p$ reference data are not available. In addition, the cross section at mid-rapidity and transverse momentum distributions in $p+p$ collisions at $\sqrt{s}$ = 39 and 62.4 GeV are evaluated. 

\end{abstract}
\pacs{}
\maketitle
\section{Introduction}
Lattice QCD predicts that, under conditions of extremely high temperatures and energy densities, a phase transition or crossover from hadronic matter to a new form of matter, known as Quark Gluon Plasma (QGP) ~\cite{QGP_bib}, will occur. The Relativistic Heavy Ion Collider (RHIC) was built to search for the QGP and to study its properties in laboratory through high-energy heavy-ion collisions~\cite{starwhitepaper}. Many observables have been proposed to probe the QGP created in heavy-ion collisions. Among them, the $J/\psi$ suppression caused by the color-charge screening  in QGP is one of most important signatures ~\cite{QGP_jpsi}.
   
Over the past twenty years, $J/\psi$ production in hot and dense medium has been a topic attracting growing interest. Suppression of $J/\psi$ production has been observed in various experimental measurements~\cite{SPS_jpsi1,SPS_jpsi2,SPS_jpsi3,PHENIX_jpsi1}. A similar suppression pattern and magnitude of $J/\psi$ was observed at SPS and RHIC despite of huge collision energy difference. Furthermore, the $J/\psi$ is suppressed more in forward rapidity than that in midrapidity at RHIC 200 GeV Au$+$Au collisions ~\cite{PHENIX_jpsi2}. These experimental observations suggest that, in addition to color screening, there exist other effects  contributing to the modification of $J/\psi$ production. Cold nuclear matter (CNM) effects, the combined contribution of finite $J/\psi$ formation time and finite space-time extent of QGP and recombination from uncorrelated c and $\bar{c}$ in the medium may account for these contributions ~\cite{10_zebo}. Among these contributions, the regeneration of $J/\psi$ from the recombination of c$\bar{c}$ plays an important role  to explain the similar suppressions at SPS and RHIC. As the collision energy increases, the regeneration of $J/\psi$ from the larger charm quark density would also increase which partly compensates for the additional suppression from color-screening. The regeneration  also expects  a stronger suppression at forward rapidity at RHIC where the charm quark density is lower than that at midrapidity. At LHC, the $J/\psi$ is less suppressed in both mid-rapidity and forward rapidity than that at RHIC ~\cite{LHC_jpsi1,LHC_jpsi2}, which may indicate that the regeneration contribution is dominant in the $J/\psi$ production at LHC energies.   Measurements of $J/\psi$  in different collision energies at the Solenoidal Tracker at RHIC (STAR)  can give us indications  on the balance of these mechanisms for $J/\psi$ production and medium properties.

To qualify the medium effects on the modification of $J/\psi$ production, the knowledge of  $J/\psi$ cross section and kinematics in $p+p$ collision  is crucial to offer a reference.  During RHIC year 2010, STAR has collected abundant events of Au$+$Au collisions at $\sqrt{s_{NN}} = 39$ and 62.4 GeV, while the reference data in $p+p$ collisions is not in the schedule of RHIC run plan. As what we did in ref ~\cite{upsilon_paper}, we study the world-wide data to obtain the $J/\psi$ reference at these collision energies.

In this letter,we report an interpolation of the $p_{T}$-integrated and differential inclusive $J/\psi$ cross section in $p+p$ collisions at mid-rapidity to $\sqrt{s} = $ 39 and 62.4 GeV. We establish a strategy to estimate the inclusive $J/\psi$ cross section and kinematics at certain energy points, which makes the calculation of the $J/\psi$ nuclear modification factors for any colliding system and energy at RHIC possible. The extrapolation is done in three steps. The first step is an energy interpolation of the existing total $J/\psi$ cross section measurements. The second step is the description of the energy evolution of the rapidity distribution. The last step is the evaluation of the energy evolution of the transverse momentum distribution. 
\section{Available experimental results treatment}
The measurements of $J/\psi$ hadroproduction have been performed for about forty years. In such a long period, different experimental techniques have been utilized and different input information was available at the time of the measurements. Therefore, comparison of different experimental results on an equal footing needs an update of the published values on several common assumptions and aspects. For example, the branching ratio of $J/\psi \rightarrow e^{+} e^{-}$ (or $\mu^{+} \mu^{-}$) have changed with time; the assumed functional forms for the $x_{F}$ and $p_{T}$ shapes, which can be used to infer the total $J/\psi$ production, are different in different measurements; and the treatment of the nuclear effects are not homogeneous. In this section, we update all the results with the current best knowledge of branching ratios, kinematics and nuclear effects.

The cross section for $J/\psi$ on a nuclear target is often characterized by a power law: 
\begin{equation}
\label{eq1}
\sigma^{pA}_{J/\psi} = \sigma_{J/\psi}^{pN} \times A^{\alpha} .
\end{equation} 
where $\sigma_{pN}^{J/\psi}$ is the $J/\psi$ proton-nucleon cross section and $\sigma_{pA}^{J/\psi}$ is the corresponding proton-nucleus cross section for a target of atomic mass number $A$. 

\renewcommand{\floatpagefraction}{0.75}
\begin{figure}[htbp]
\begin{center}
\includegraphics[keepaspectratio,width=0.45\textwidth]{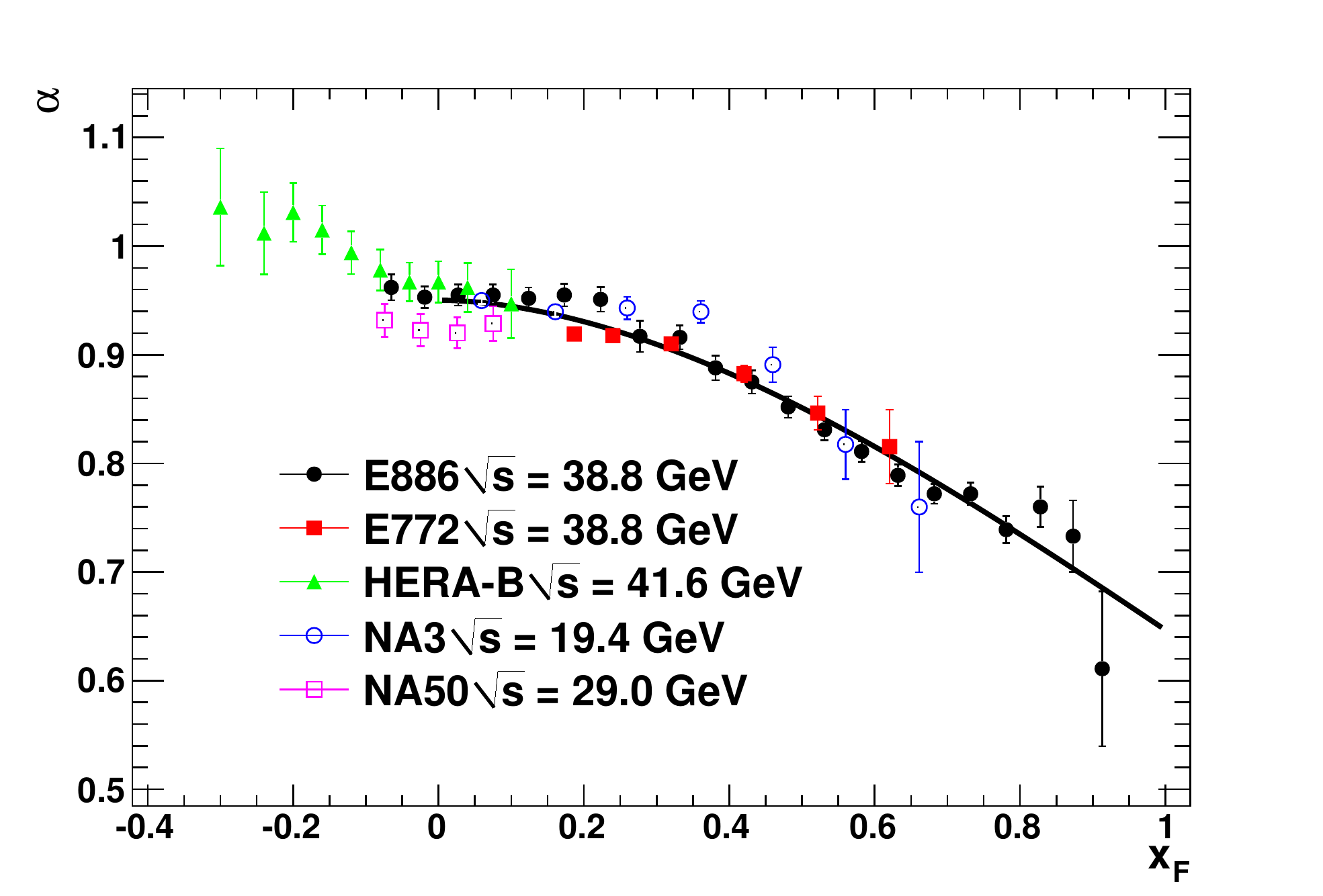}
\caption{(color online) Measurements of $\alpha$ as a function of $x_{F}$ by various experiments in different collision energies. The solid curve represents the parametrization discussed in the text.}
\label{figure1}
\end{center}
\end{figure}
The dependence of $\alpha$ on $x_{F}$ measured by NA3 ~\cite{alpha_na3}, NA50 ~\cite{alpha_na50}, E772 ~\cite{alpha_E772}, E886 ~\cite{alpha_E886} and HERA-B ~\cite{alpha_HERA-B} is shown in Fig.~\ref{figure1}, where $x_{F}$ is defined as $x_{F} = 2p_{z}/\sqrt{s}$ ($p_{z}$ is longitudinal momentum, along the beam direction.). No significant energy dependence of $\alpha$ as a function of $x_{F}$ is observed within uncertainties, thus we assume it is independent of the cms-energy ($\sqrt{s}$). The results of $J/\psi$ $\alpha$ at $x_{F}>0$ can be represented for convenience by the simple parametrization shown as solid line in Fig.~\ref{figure1}:    $\alpha(x_{F}) =0.9503e^{-ln2(\frac{x_{F}}{1.3846})^{1.8067}} $ . If an experiment has published cross sections of $J/\psi$ in proton nucleus collisions, Eq.~\ref{eq1} and the solid curve in Fig.~\ref{figure1} is applied to obtain the corresponding $J/\psi$ cross section in proton nucleon collisions. Some of the experimental measurements are only quoted for a limited phase-space. To obtain the total cross sections, the functional forms of $x_{F}$ and $p_{T}$ spectrum shapes utilized for extrapolation are: $d\sigma/dx_{F} = a \times e^{-ln2(\frac{x_{F}}{b})^{c}}$,and $d\sigma/dp_{T} = a \times \frac{p_{T}}{(1+b^{2}p_{T}^{2})^{c}}$ respectively, where a, b, and c are free parameters. As illustrated in Fig.~\ref{E331_Fig}, these two functional forms describe the $x_{F}$ and $p_{T}$ spectra very well.  All the measurements are updated with the latest branching fractions ($5.961 \pm 0.032\%$ for $J/\psi \rightarrow \mu^{+} + \mu^{-}$, $5.971 \pm 0.032\%$ for $J/\psi \rightarrow e^{+}+e^{-}$) ~\cite{PDG}. The treated results on $J/\psi$ cross sections are listed in Tab.~\ref{table1}.They show a good overall consistency, even though some of them contradict with each other. For example, the two measurements (E331 and E444) at 20.6 GeV deviate from each other by roughly 2$\sigma$. The E705 measurement at 23.8 GeV is higher than the UA6 one at 24.3 GeV by more than 2$\sigma$.

\renewcommand{\floatpagefraction}{0.75}
\begin{figure*}[htbp]
\begin{center}
\includegraphics[keepaspectratio,width=0.45\textwidth]{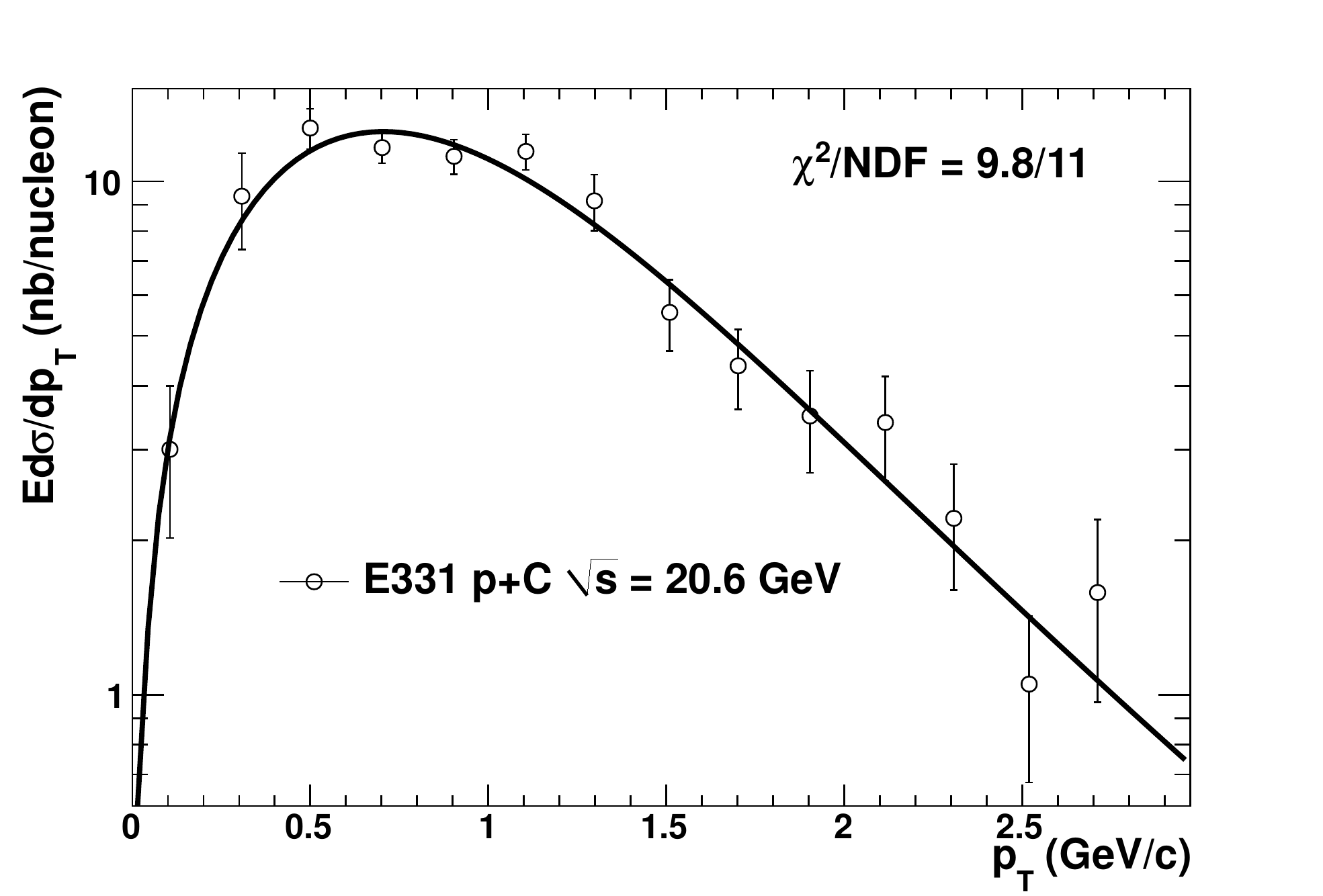}
\includegraphics[keepaspectratio,width=0.45\textwidth]{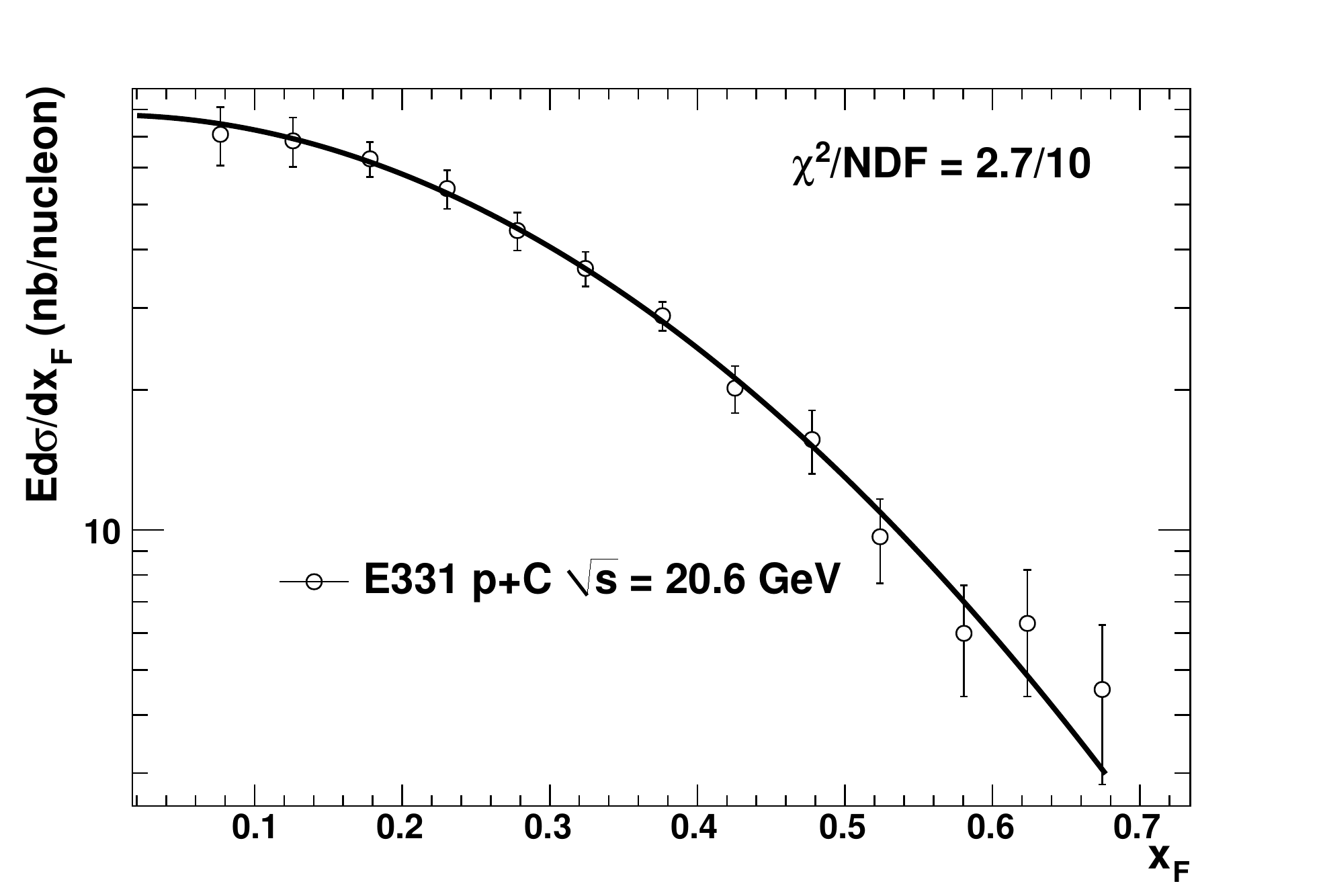}
\caption{(color online) Distributions of  $Ed\sigma/dp_{T}$ (left panel) and $Ed\sigma/dx_{F}$ (right panel) in $p+C$ collisions at $\sqrt{s} = 20.6$ GeV measured by E331 collaboration ~\cite{E331_1}. The solid lines are fit curves with the functional forms described in the text.}
\label{E331_Fig}
\end{center}
\end{figure*}

\renewcommand{\floatpagefraction}{0.75}
\begin{table*}[htbp]
\newcommand{\tabincell}
\centering
\begin{center}
\begin{tabular}{|c|c|c|c|c|c|c|c|}
\hline
Experiment&Reaction&$\sqrt{s}$ (GeV)&$\sigma_{J/\psi}$ (nb/nucleon)\\
\hline
\hline
CERN-PS ~\cite{CERN-PS}&p+A &6.8&0.732$\pm$0.13\\
\hline
WA39 ~\cite{WA39}&p+p&8.7&2.35$\pm$1.18\\
\hline
IHEP ~\cite{IHEP}&p+Be&11.5&21.63$\pm$5.64\\
\hline
E331 ~\cite{E331}&p+Be&16.8&85.15$\pm$21.30\\
\hline
NA3 ~\cite{alpha_na3}&p+Pt&16.8&95.0$\pm$17.0\\
\hline
NA3 ~\cite{alpha_na3}&p+Pt&19.4&122.6$\pm$21\\
\hline
NA3 ~\cite{alpha_na3}&p+p&19.4&120$\pm$22\\
\hline
E331 ~\cite{E331_1}&p+C&20.6&278$\pm$32.8\\
\hline
E444 ~\cite{E444}&p+C&20.6&176.5$\pm$23.3\\
\hline
E705 ~\cite{E705}&p+Li&23.8&271.51$\pm$29.84\\
\hline
UA6 ~\cite{UA6}&p+p&24.3&171.42$\pm$22.21\\
\hline
E288 ~\cite{E288}&p+Be&27.4&294.12$\pm$73.53\\
\hline
E595 ~\cite{E595}&p+Fe&27.4&264$\pm$56\\
\hline
NA38/51 ~\cite{NA38, NA51}&p+A&29.1&229.5$\pm$34.4\\
\hline
NA50 ~\cite{alpha_na50}&p+A&29.1&250.7$\pm$37.6\\
\hline
E672/706 ~\cite{E672}&pBe&31.6&343.07$\pm$75.12\\
\hline
E771 ~\cite{E771}&p+Si&38.8&359.1$\pm$34.2\\
\hline
E789 ~\cite{E789}&p+Au&38.8&415.04$\pm$100\\
\hline
ISR ~\cite{ISR52}&p+p&52&716$\pm$303\\
\hline
PHENIX ~\cite{PHENIX200}&p+p&200&4000$\pm$938\\
\hline
CDF ~\cite{CDF1960}&p+$\bar{p}$&1960&22560$\pm$3384\\
\hline
ALICE ~\cite{ALICE2760}&p+p&2760&29912.6$\pm$5384.3\\
\hline
ALICE ~\cite{ALICE7000}&p+p&7000&54449.4$\pm$8494\\
\hline
\end{tabular}
\end{center}
\caption{(color online) Updated total ($\sigma_{J/\psi}$) production cross sections in proton-induced interactions.}
\label{table1}
\end{table*} 
\section{Results}
\renewcommand{\floatpagefraction}{0.75}
\begin{figure}[htbp]
\begin{center}
\includegraphics[keepaspectratio,width=0.45\textwidth]{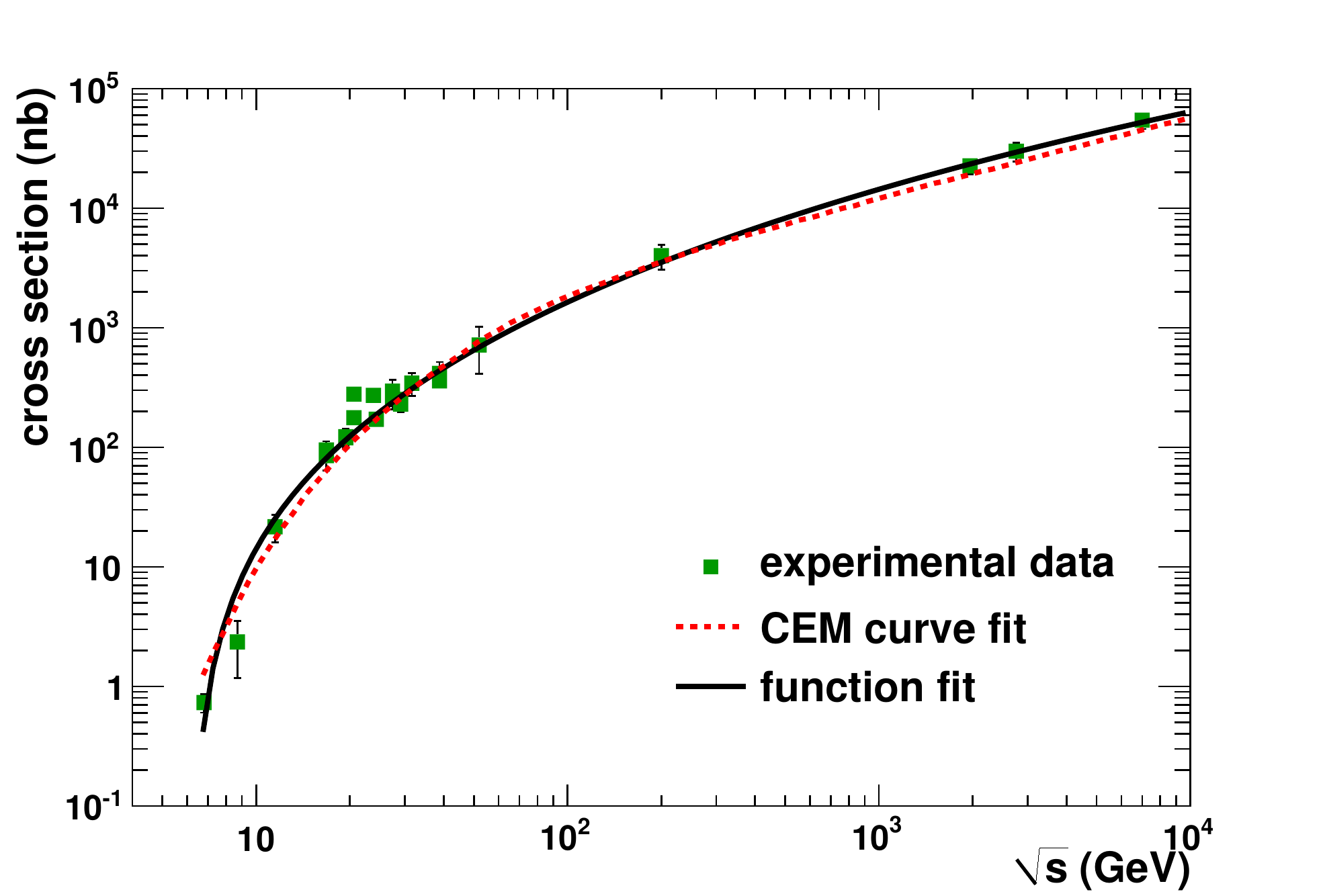}
\caption{(color online) Energy dependence of inclusive $J/\psi$ prodcution cross section. The open circle is the fit from CEM shape.The solid line is a function fit as discuss in the text.}
\label{figure2}
\end{center}
\end{figure}
The energy evolution of the total inclusive $J/\psi$ production cross section in proton induced interactions is shown in Fig.~\ref{figure2}. The first approach is to use the predicted shape in the Colour Evaporation Model at Next to Leading Order (NLO) ~\cite{CEM} to describe the energy dependence of $J/\psi$ cross section. The central CT10 parton density set ~\cite{PDF_CT10} and $\{m,\mu_{F}/m,\mu_{R}/m\}=\{1.27 (GeV), 2.10, 1.60\}$ set is utilized in the predicted shape, where m is the charm quark mass, $\mu_{F}$ is the factorization scale, $\mu_{R}$ is the renormalization scale. The fit is defined such that the normalization of the NLO CEM calculation is left as a free parameter ($\alpha$): $\sigma=\alpha \times \sigma_{CEM}$. The second approach is to use a functional form to describe the cross section energy evolution: $f(\sqrt{s})=a \times y_{max}^{d}\times e^{\frac{-b}{y_{max}+c}}$, where $y_{max} = ln(\frac{\sqrt{s}}{m_{J/\psi}})$, a, b, c and d are free parameters. As shown in Fig.~\ref{figure2}, both approaches can describe the energy evolution trend of $J/\psi$ cross section. The $\chi^{2}/NDF$ for CEM and functional fit are 90.5/22 and 76.7/20, respectively. The large $\chi^{2}$ mainly comes from three experimental points which contradict with the common trend (E331 and E444 measurements at 20.6 GeV, E705 measurement at 23.8 GeV). If we exclude these three data points and refit the results, the  $\chi^{2}/NDF$ for CEM and functional fit are 41.1/19 and 16.7/16, respectively. The values extrapolated (without the three bad experimental points) for the $J/\psi$ cross sections at $\sqrt{s} = $ 39 and 62.4 GeV, utilizing the functional form and the NLO CEM based fit are listed in Table~\ref{table2}.
\renewcommand{\floatpagefraction}{0.75}
\begin{table}[htbp]
\newcommand{\tabincell}
\centering
\begin{center}
\begin{tabular}{|c|c|c|}
\hline
\multirow{2}{*}{Fit} & \multicolumn{2}{|c|}{cross section (nb/nucleon)}\\\cline{2-3}&$\sqrt{s} = $39 GeV &$\sqrt{s} = $62.4 GeV\\
\hline
NLO CEM& 425$\pm$20&941$\pm$42\\
\hline
function&445$\pm$30&995$\pm$66\\
\hline
evaluated results&445$\pm$30$\pm$20&995$\pm$66$\pm$54\\
\hline
\end{tabular}
\end{center}
\caption{Extrapolated values of the $J/\psi$ production cross section at $\sqrt{s} =$ 39 and 62.4 GeV. The difference between CEM and function fit has been taken as the systematic uncertainties of the extrapolation.}
\label{table2}
\end{table} 

\renewcommand{\floatpagefraction}{0.75}
\begin{figure}[htbp]
\begin{center}
\includegraphics[keepaspectratio,width=0.45\textwidth]{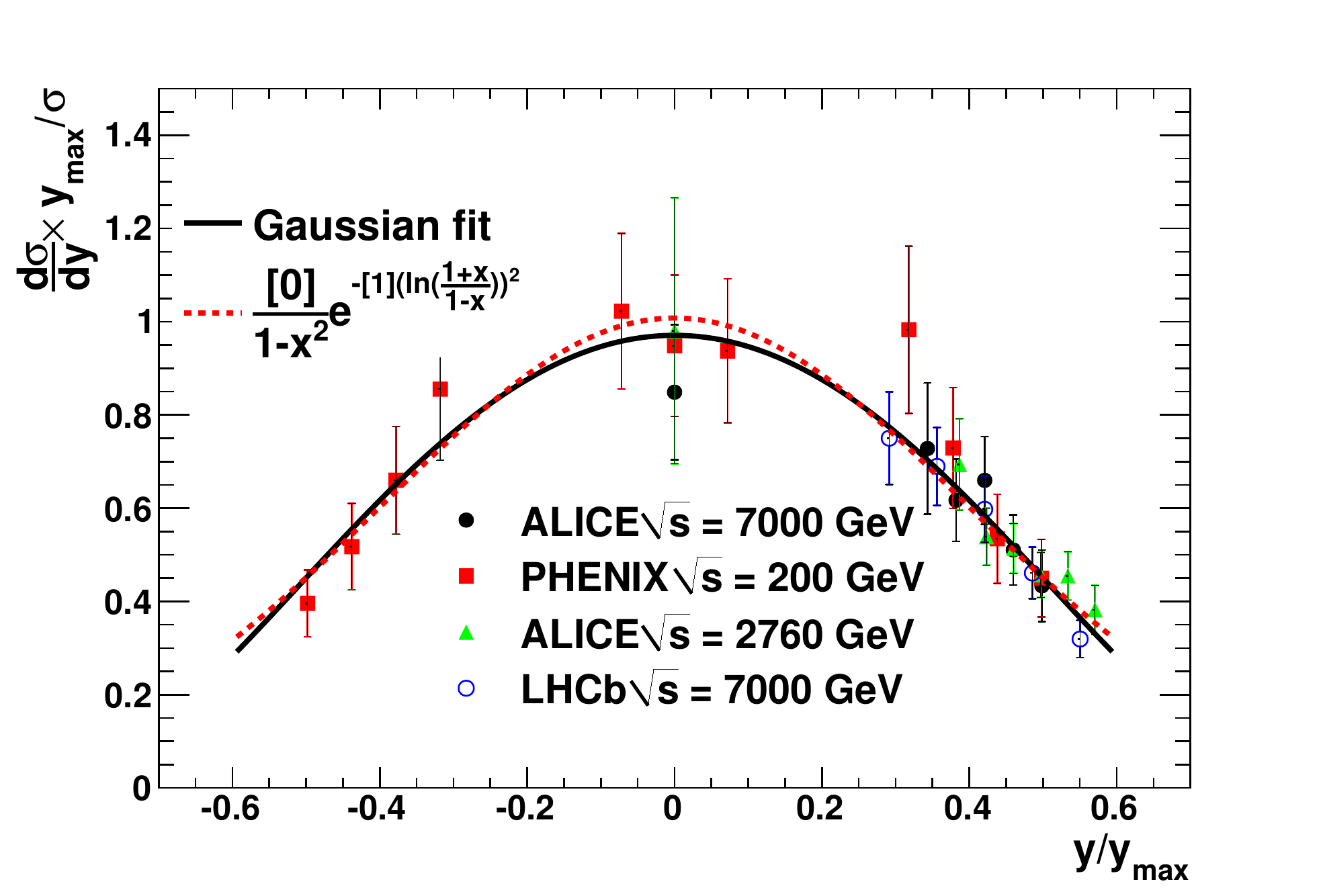}
\caption{(color online) Normalized $J/\psi$ production cross section as a function of $y/y_{max}$. Two function fits are shown: one is Gaussian function, the other one is $\frac{a}{1-(y/y_{max})^{2}}e^{-b(ln(\frac{1+y/y_{max}}{1-y/y_{max}}))^{2}}$. The difference between these two curves has been considered as systematic errors.}
\label{figure3}
\end{center}
\end{figure}

The knowledge of the rapidity dependence of $J/\psi$ production at different cms-energies is crucial to obtain a reference for the measurements at mid-rapidity from RHIC. Based on a universal energy scaling behavior in the rapidity distribution obtained at different cms-energies, we explore approaches to the extrapolation of the rapidity distribution. As shown in Fig.~\ref{figure3}, the y-differential cross sections at different cms-energies have been normalized by the total cross section, and the normalized values are plotted verse $y/y_{max}$, where $y_{max}$ has been previously defined. Despite of huge cms-energy difference, the treated RHIC ~\cite{PHENIX200} and LHC ~\cite{ALICE2760, ALICE7000, LHCb7000} experimental distributions fall into a universal trend, which allows us to perform global fits to all the experimental results with suitable functions. Two functional forms are chosen to do the fits: one is Gaussian function, the other one is  $\frac{a}{1-(y/y_{max})^{2}}e^{-b(ln(\frac{1+y/y_{max}}{1-y/y_{max}}))^{2}}$, where a and b are free parameters. Both of them can describe the global distribution very well($\chi^{2}/NDF = 10.1/27$ for gaussian fit, $\chi^{2}/NDF = 11.2/27$ for the other fit). With the extrapolated $J/\psi$ cross sections and rapidity distributions, the predicted $J/\psi$ cross section times branching ratio at $\sqrt{s} =$ 39 and 62.4 GeV in mid-rapidity are $Br(e^{+}e^{-})d\sigma/dy|_{|y|<1.0} = 9.04 \pm 0.69$ and $17.74 \pm 1.06$ nb, respectively.  The uncertainties include statistical and systematic uncertainties. These values are highly consistent with the estimations from CEM model ($8.7 \pm 4.5$ nb for 39 GeV, $17.4 \pm 8.0$ for 62.4 GeV).

\renewcommand{\floatpagefraction}{0.75}
\begin{figure}[htbp]
\begin{center}
\includegraphics[keepaspectratio,width=0.45\textwidth]{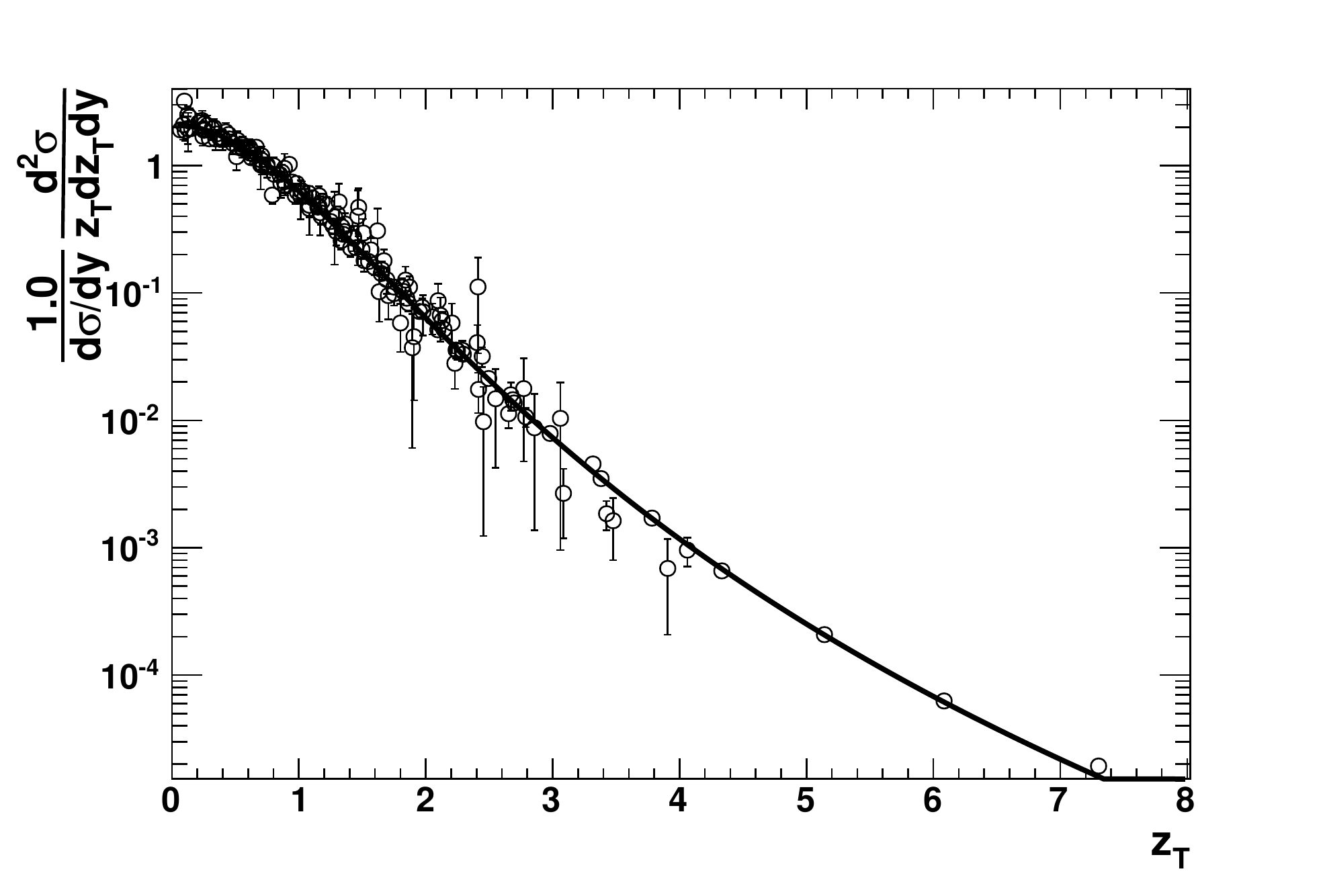}
\caption{(color online) $J/\psi$ $z_{T}$ distributions  for available experimental results  from $\sqrt{s} =$ 10 to 7000 GeV. The solid line is a function fit as discussed in the text.}
\label{figure4}
\end{center}
\end{figure}

The energy evolution of $J/\psi$ transverse momentum distribution are also studied via available experimental measurements from $\sqrt{s} = $ 10 - 7000 GeV ~\cite{alpha_na3, E288, E672, E331_1,PHENIX200, ALICE7000, CDF1960, ISR53}. We use part of the world-wide fixed-target data (with only  p, Be, Li, and C respectively) measured with incident protons. In this way, we avoid uncertainties due to ignoring any cold nuclear matter effects on the $J/\psi$ transverse momentum distributions. To compare the different experimental measurements at different energies and rapidity domains, as shown in Fig.~\ref{figure4}, the transverse momentum distributions are normalized by their $p_{T}$-integrated cross sections and plotted verse the $z_{T}$ variable, which is defined as $z_{T} = p_{T}/<p_{T}>$.  The treated distributions follow a universal trend despite of the different cms-energies and rapidity domains. We can describe the global distributions very well by the following function: $\frac{1}{d\sigma/dy}\frac{d^{2}\sigma}{z_{T}dz_{T}dy} =a \times \frac{1}{(1+b^{2}z_{T}^{2})^{n}}$ ~\cite{function}, where $a=2b^{2}(n-1)$, $b=\Gamma(3/2)\Gamma(n-3/2)/\Gamma(n-1)$, and n is the only free parameter.  From the fit, we obtain $n=3.93 \pm 0.03$ with $\chi^{2}/NDF = 143.9/162$. 

\renewcommand{\floatpagefraction}{0.75}
\begin{figure}[htbp]
\begin{center}
\includegraphics[keepaspectratio,width=0.45\textwidth]{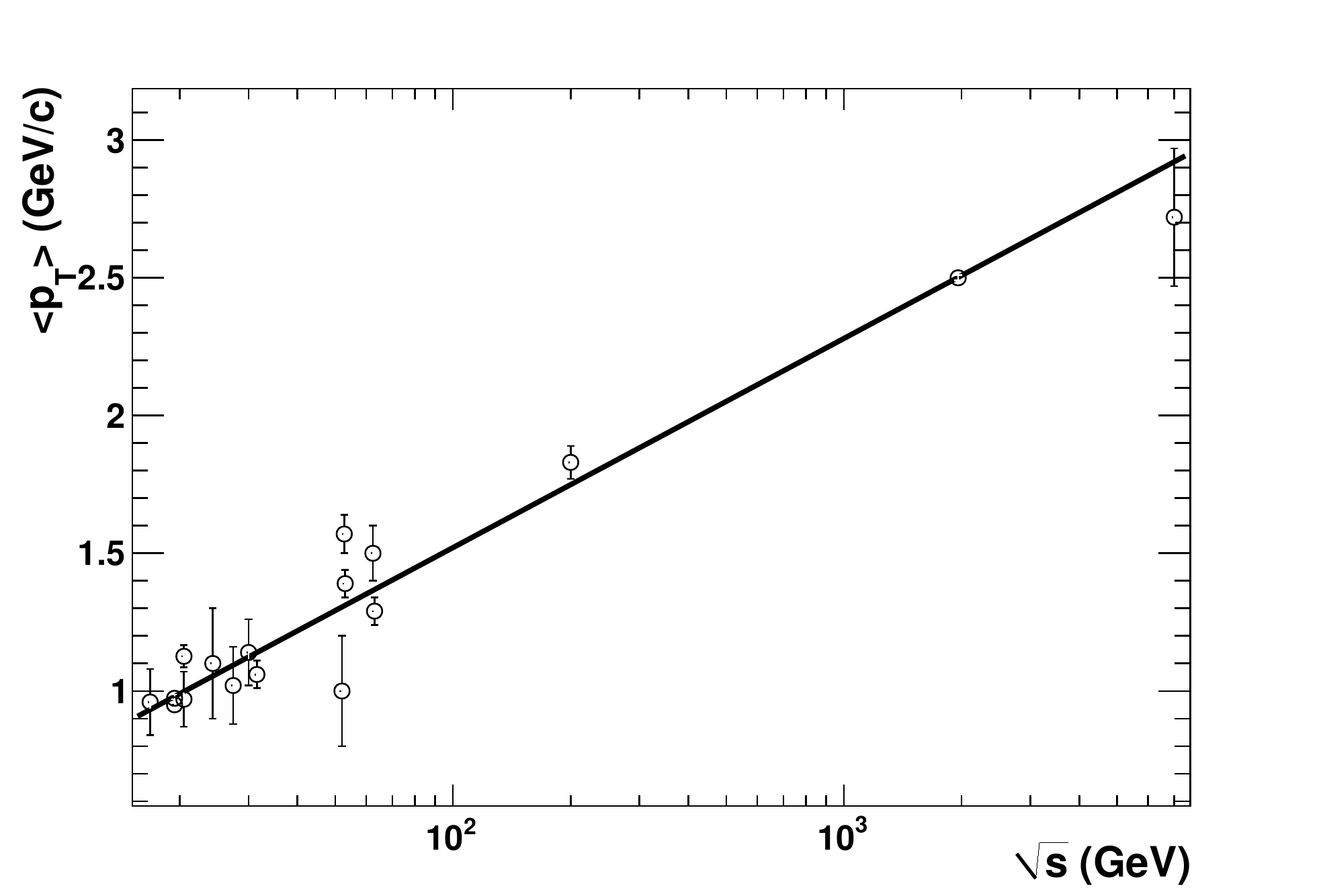}
\caption{(color online) $J/\psi$ $<p_{T}>$ at mid-rapidity as a function of cms-energy  from $\sqrt{s} =$ 10 to 7000 GeV. The solid line is a function fit as discussed in the text.}
\label{figure5}
\end{center}
\end{figure}
With the universal shape and $<p_{T}>$ information at certain energy and rapidity domain (we focus on mid-rapidity), we can extrapolate the transverse momentum distribution at any cms-energy. Thus the next step is to evaluate the energy evolution of $<p_{T}>$. The $<p_{T}>$ at mid-rapidity as a function of cms-energy from world-wide experiments ~\cite{alpha_na3, E288, E672, E331_1,PHENIX200, ALICE7000, CDF1960, ISR53, ISR30, ISR52}  is shown in Fig.~\ref{figure5}. Also, only part of the world-wide fixed-target data (with p, Be, Li, and C respectively) are used to reduce the cold nuclear matter effects. The $<p_{T}>$ versus energy can be fitted by the function form: $f(\sqrt{s}) = p + q ln\sqrt{s}$, where p, q are free parameters. The fit parameters are $p=0.0023 \pm 0.0182$, $q = 0.329 \pm 0.031$ with $\chi^{2}/NDF = 41.1/15$.  The estimated $<p_{T}>$ from the fit function at $\sqrt{s} =$ 39 and 62.4 GeV are $1.21 \pm 0.04$ and $1.36 \pm 0.04$ GeV/c, respectively. With these inputs, the transverse momentum distribution at these two cms-energies can be completely determined.  

\renewcommand{\floatpagefraction}{0.75}
\begin{figure}[htbp]
\begin{center}
\includegraphics[keepaspectratio,width=0.45\textwidth]{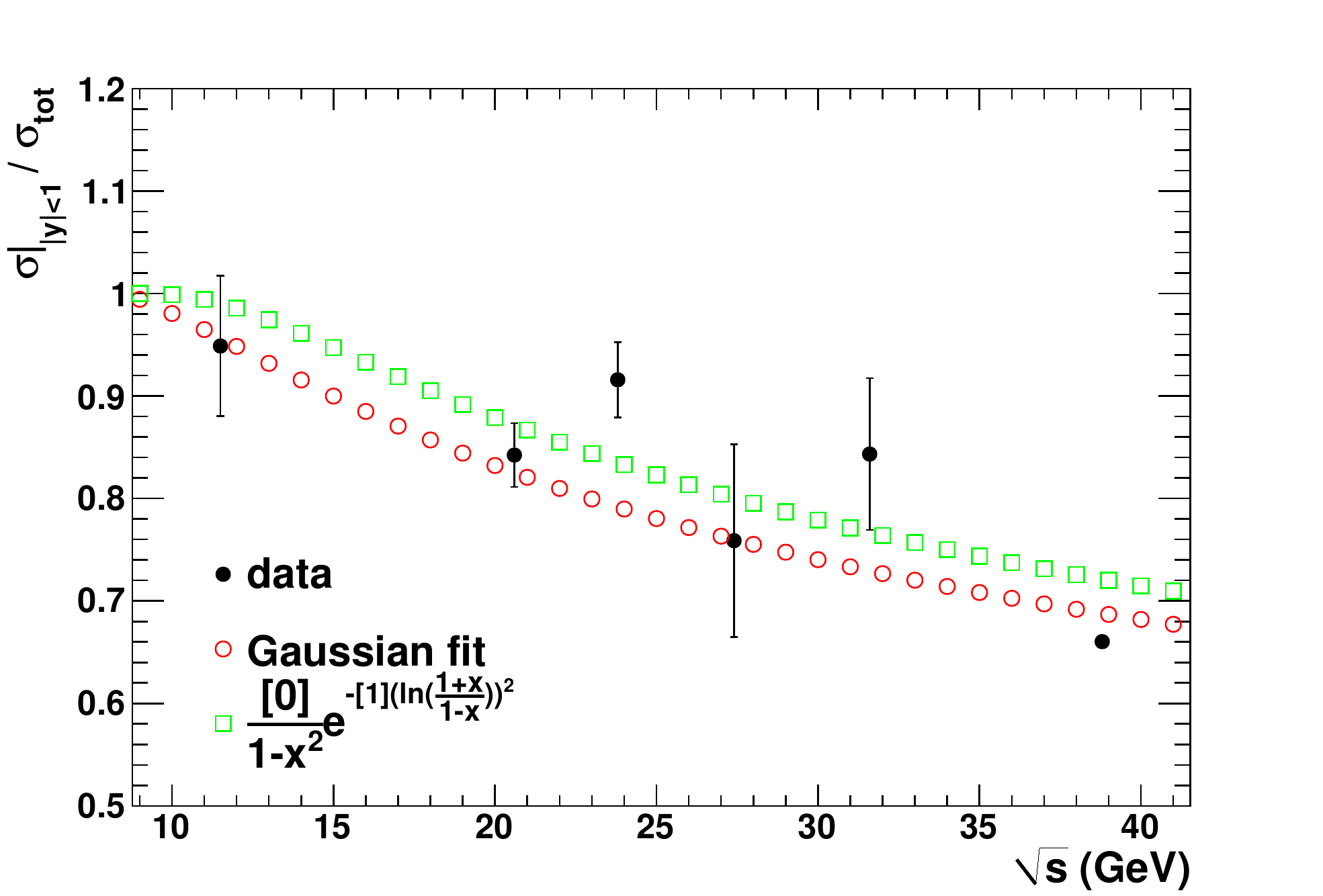}
\caption{(color online) The ratios of  $J/\psi$ $\sigma|_{|y|<1.0}$ to $\sigma_{total}$ as a function of  cms-energy. The open points are the estimations using the two fit functions in Fig.~\ref{figure3}.}
\label{figure6}
\end{center}
\end{figure}

There are rare  rapidity distribution measurements in p$+$A collisions at $\sqrt{s} < 200$. Therefore, the universal energy scaling parameters of rapidity distributions are determined by the measurements at $\sqrt{s} \geq 200$ GeV.  Its validity at low energy ($<$200 GeV) range still need to be further investigated. But we do have various  $x_{F}$ distribution measurements of  $J/\psi$ in fix-target experiments ~\cite{IHEP, E331_1, E444, E595, E705, E672, E771, E789}. 
Cooperated with the $\alpha$ verse $x_{F}$ curve in Fig.~\ref{figure1} and the transverse momentum distributions obtained using the strategy described in the above section, we can evaluate the rapidity distributions via the $x_{F}$ distributions measurements in the fix-target experiments to check the validity of the rapidity interpolation method.  The ratios of  $J/\psi$ $\sigma|_{|y|<1.0}$ to $\sigma_{total}$, which are calculated utilizing the evaluated rapidity distributions in fix-target experiments, verse cms-energy are shown in Fig.~\ref{figure6}. The open points plotted in the figure are the estimations using the two fit functions in Fig.~\ref{figure3}.  In this figure, we can see that our extrapolation strategy also works at low cms-energy range. 
\section{summary} 
We study the world-wide data of $J/\psi$ production and kinematics  at  $\sqrt{s} = 6.8-7000$ GeV. We have developed a strategy to interpolate the $J/\psi$ cross section, rapidity distribution, and transverse momentum distribution at any cms-energy in  $\sqrt{s} = 6.8-7000$ GeV. The rapidity and transverse momentum distributions measured in different energies have a universal energy scaling behavior. With this strategy, we predicted that the $J/\psi$ cross section times branching ratio at $\sqrt{s} =$ 39 and 62.4 GeV in mid-rapidity are $Br(e^{+}e^{-})d\sigma/dy|_{|y|<1.0} = 9.04 \pm 0.69$, $17.74 \pm 1.06$ nb, respectively.

\section{Acknowledgments}
We express our gratitude to the STAR Collaboration and the RCF at BNL for their support. This work was supported in part by the U.S. DOE Office of Science under the contract No. DE-SC0012704; authors Wangmei Zha and Chi Yang are supported in part by the National Natural Science Foundation of China under Grant Nos 11005103 and 11005104, China Postdoctoral Science Foundation funded project, and the Fundamental Research Funds for the Central Universities.


\begin{thebibliography}{9}
\bibitem{QGP_bib} P. Braun-Munzinger, J. Stachel,  Nature \textbf{448}, 302 (2007).
\bibitem{starwhitepaper} J. Adams et al. (STAR Collab.), Nucl. Phys. A \textbf{757}, 102 (2005).
\bibitem{QGP_jpsi} T. Matsui, H. Satz,  Phys. Lett. B \textbf{178}, 416 (1986).
\bibitem{SPS_jpsi1} M.C. Abreu et al., Phys. Lett. B \textbf{477}, 28 (2000).
\bibitem{SPS_jpsi2} B. Alessandro et al., Eur. Phys. J. C \textbf{39}, 335 (2005).
\bibitem{SPS_jpsi3} B. Alessandro et al., Eur. Phys. J. C \textbf{48}, 329 (2006).
\bibitem{PHENIX_jpsi1}  A. Adare et al., Phys. Rev. Lett. \textbf{98}, 232301 (2007).
\bibitem{PHENIX_jpsi2} A. Adare et al., Phys. Rev. C \textbf{84}, 054912 (2011).
\bibitem{10_zebo}F. Karsch et al. Phys. Lett. B \textbf{193}, 105 (1987)
\bibitem{LHC_jpsi1} B. Abelev et al., Phys. Rev. Lett. \textbf{109}, 072301 (2012).
\bibitem{LHC_jpsi2} B. Abelev et al., Phys. Lett. B \textbf{734}, 314 (2014).
\bibitem{upsilon_paper} Phys. Rev. C \textbf{88}, 067901 (2013).
\bibitem{alpha_na3} J. Badier et al., Z. Phys. C \textbf{20}, 101 (1983).
\bibitem{alpha_na50} B. Alessandro et al. (NA50 Collab.), Eur. Phys. J. C \textbf{33}, 31 (2004).
\bibitem{alpha_E772} D. M. Alde et al., Phys. Rev. Lett. \textbf{72}, 1318 (1994).
\bibitem{alpha_E886} M. J. Leitch et al., Phys. Rev. Lett.\textbf{84}, 3256 (2000).
\bibitem{alpha_HERA-B} I. Abt et al., Eur. Phys. J. C \textbf{60}, 525 (2009).
\bibitem{PDG} K. A. Olive et al., Chin. Phys. C \textbf{38}, 090001 (2014).
\bibitem{CERN-PS} A. Bamberger et al., Nucl. Phys. B \textbf{134}, 1 (1978).
\bibitem{WA39} M. J. Corden et al., Phys. Lett. B \textbf{98}, 220 (1981).
\bibitem{IHEP} Yu. M. antipov et al., Phys. Lett. B \textbf{60}, 309 (1976).
\bibitem{E331} K. J. Anderson et al., Phys. Rev. Lett. \textbf{36}, 237 (1976).
\bibitem{E331_1} J. G. Branson et al., Phys. Rev. Lett. \textbf{38}, 1331 (1977).
\bibitem{E444} K. J. Anderson et al., Phys. Rev. Lett. \textbf{42}, 944 (1979).
\bibitem{E705} L. Antoniazzi et al., Phys. Rev. D \textbf{46}, 4828 (1992).
\bibitem{UA6} C. Morel et al., Phys. Lett. B \textbf{252}, 505 (1990).
\bibitem{E288} H. D. Snyder et al., Phys. Rev. Lett. \textbf{36}, 1415 (1976).
\bibitem{E595} E. J. Siskind et al., Phys. Rev. D \textbf{21}, 628 (1980).
\bibitem{E705} L. Antoniazzi et al., Phys. Rev. D \textbf{46}, 4828 (1992).
\bibitem{NA38} M. C. Abreu et al., Phys. Lett. B \textbf{444}, 516 (1998).
\bibitem{NA51} M. C. Abreu et al., Phys. Lett. B \textbf{438}, 35 (1998).
\bibitem{E672} A. Gribushin et al., Phys. Rev. D \textbf{62}, 012001 (2000).
\bibitem{E771} T. Alexopoulos et al., Phys. Rev. D \textbf{55}, 3927 (1997).
\bibitem{E789} M. H. Schub et al., Phys. Rev. D \textbf{52}, 1307 (1995).
\bibitem{ISR52} E. Nagy et al., Phys. Lett. B \textbf{60}, 96 (1975).
\bibitem{PHENIX200} A. Adare et al. (PHENIX Collab.), Phys. Rev. Lett. \textbf{98}, 232002 (2007).
\bibitem{CDF1960} D. Acosta et al. (CDF Collab.), Phys. Rev. D \textbf{71}, 032001 (2005).
\bibitem{ALICE2760} B. Abelev et al. (ALICE Collab.), Phys. Lett. B \textbf{718}, 295 (2012).
\bibitem{ALICE7000} K. Aamodt et al. (ALICE Collab.), Phys. Lett. B \textbf{704}, 442 (2011).
\bibitem{CEM} R. E. Nelson, R. Vogt, and A. D. Frawley et al., Phys. Rev. C \textbf{87}, 014908 (2013).
\bibitem{PDF_CT10} H. L. Lai et al., Phys. Rev. D \textbf{82}, 074024 (2010).
\bibitem{LHCb7000} R. Aaij et al. (LHCb Collab.), Eur. Phys. J. C \textbf{71}, 1645 (2011). 
\bibitem{function} F. Bossu et al., arXiv:1103.2394 [nucl-ex].
\bibitem{ISR53} B. Aubert et al. , Nucl. Phys. B \textbf{1421}, 29 (1978). 
\bibitem{ISR30} E. Nagy et al., Phys. Lett. B \textbf{60}, 96 (1975). 
\bibitem{ISR52} E. Amaldi et al., Nuovo Cim. \textbf{19}, 152 (1977). 
\end{thebibliography}
\end{document}